# Site-Resolved Learning of Reticular Frameworks with Equivariant Neural Networks

Zhenhao Zhou[1,2], Salman Bin Kashif[3], Han Hao[4], Jin-Hu Dou[5], Chris Wolverton[6], Kaihang Shi[3]*, Tao Deng[1], Zhenpeng Yao[1,2,4]*

[1]Center of Hydrogen Science and School of Materials Science and Engineering, Shanghai Jiao Tong University, Shanghai 200240, China

[2]Innovation Center for Future Materials, Zhangjiang Institute for Advanced Study, Shanghai Jiao Tong University, Shanghai 201203, China

[3]Department of Chemical and Biological Engineering, University at Buffalo, The State University of New York, Buffalo, New York 14260, United States

[4]Acceleration Consortium, 700 University Avenue, Toronto, Ontario M7A 2S4, Canada

[5]National Key Laboratory of Advanced Micro and Nano Manufacture Technology and Key Laboratory of Polymer Chemistry and Physics of Ministry of Education, School of Materials Science and Engineering, Peking University, Beijing, Beijing 100871, China

[6]Department of Materials Science and Engineering, Northwestern University, Evanston, Illinois 60208, USA

*E-mail: kaihangs@buffalo.edu; yaozhenpeng@gmail.com.

**Abstract:**

Reticular frameworks hold promise for diverse sustainable applications, yet their immense chemical space limits efficient and systematic design. While machine learning provides a compelling pathway to accelerate exploration, existing approaches often lack either interpretability or fidelity in linking crystal geometry to emergent properties. Here, we introduce a site-resolved equivariant learning framework based on three-dimensional periodic space sampling, which decomposes reticular structures into local geometric environments for simultaneous property prediction and site-wise contribution analysis. Trained on a combination of constructed and retrieved datasets, the model achieves state-of-the-art accuracy and data efficiency across gas storage, gas separation, and electronic-property prediction tasks. Importantly, the framework reveals interpretable local structure-property relationships by identifying transferable high-contribution sites across diverse frameworks. Leveraging these learned motifs, we further demonstrate inverse design of new metal-organic frameworks exhibiting record-high $N_2$ storage, strong $CO_2/N_2$ separation performance, and near-zero electronic band gaps, validated by physics-based simulations.

## Introduction

Reticular frameworks, featured by their characteristic pores and voids with diameters varying from several to a hundred nanometers, encompass a diverse family of structures including metal-organic frameworks (MOFs) and covalent organic frameworks (COFs). Their modular construction, high internal surface areas, and chemically tunable pore environments have made them pivotal platforms[1–4] for the development of clean energy applications and beyond, such as catalysis[5–8], gas storage and separation[9,10], advanced energy storage[11–15], sensing[16,17] , ionic conduction,[15,18,19] and drug delivery [20,21]. However, designing a reticular framework for a specific application remains challenging. In the case of MOFs, the design process of a MOF entails selecting suitable metal nodes and organic linkers to self-assemble following a specific topology, and performing a proper post-synthetic treatment.[22–24] To date, more than a hundred different metal nodes and topologies have been reported, and the number of potentially viable linkers has been estimated to be vast.[25,26] Together, these choices define an enormous chemical space for MOF, and more broadly, the reticular frameworks. An effective and efficient exploration of this space thus demands approaches beyond the Edisonian-like trial-and-error method, and data-driven techniques like machine learning offer a promising way.

Machine-learning workflows in materials science seek to connect a material's composition and structure with its resulting properties, which generally starts by converting each material to a numerical representation that a machine learning model can learn from and predict.[27,28] As a result, ML models are often dictated by the choice of representation; for instance, a graph-based encoding naturally leads to a graph neural network[29], whereas a voxelized or image-like representation calls for a convolutional architecture[30]. Just as with other crystalline materials, researchers have developed a variety of representations specifically for reticular frameworks, each designed to capture their chemical, physical, and crystallographic information. Descriptor-based representations, such as stoichiometric features[31] and smooth overlap of atomic positions (SOAP)[31] have been proposed for reticular materials. These can help reduce input dimensionality and allow the employment of simpler models. Yet, their reliance on precomputed and partial features rather than the full structural detail often limits predictive performance.[32–34] Graph representation has also been applied, but converting the reticular structure into a conventional crystal graph will lose the information about its three-dimensional (3-D) conformation, which is critical for accurate prediction of properties like gas adsorption.[31,35]

One notable representation scheme for reticular framework is the energy grid-based representations, including the one-dimensional (1-D) energy histogram[36], the two-dimensional (2-D) energy histogram[37], and the 3-D voxel[38–40]. In the 1-D representation creation, the framework is gridded, the interaction energy between the material and the probe atom at each grid point is calculated (energy grid), and the energy distribution is then collected as the input. The 2-D variant extends this approach by incorporating energy gradient magnitudes as a second dimension. Both the 1-D and 2-D histograms have shown impressive results for achieving high accuracy with relatively simple regression methods like multilayer perceptron (MLP)[36] and LASSO[37]. The 3-D voxel representations preserve spatial correlations by directly processing raw energy grids through 3-D convolutional neural networks (CNNs), thereby retaining localized geometric features that lower-dimensional projections inherently discard.[38–40] Yet it is challenging to enforce translational and rotational symmetry with the 3-D voxels, and related models therefore require significantly more data to train[27,41]. The same host-guest energy-landscape perspective has also been extended to more task-specific applications, including potential-energy-field analysis for diffusion[42] , transition-state and geometry descriptors for diffusion prediction[43], surface sampling of adsorption enthalpy[44], Voronoi-energy descriptors for mixture-separation screening[45], and PoroNet[46] for high-throughput pore screening. These approaches are highly valuable, but they generally rely on precomputed host-guest interaction fields for a chosen adsorbate and sampling protocol. As a result, the connection from model behavior back to transferable structural motifs can be less direct than in a purely structure-based representation. Besides, hybrid embeddings (*e.g.*, RFcode) encode the chemically meaningful organic linkers as SMILES[47]/graphs[26] while representing metal nodes and topology with simplified symbolic tokens for better overall efficiency. Hybrid embeddings can be readily processed by recurrent neural network/transformer-based architectures (as demonstrated by SmVAE and MOFormer[26,31,48,49]). However, this convenience comes at the cost of discarding explicit 3-D structural information. To date, incorporating comprehensive structural information into representations for reticular materials is still an open challenge.

For ML models operating on 3-D atomic structures, it is desirable for the representation to respect the Euclidean symmetries of space[50–52]. The recent development of equivariant neural networks coupled with 3-D atom graph representations presents an elegant solution to material representation challenges, which inherently ensures these rules through tensor products on spherical harmonic functions.[52–54] Equivariant models enhance data efficiency by incorporating prior knowledge of symmetries, eliminating the need for data augmentation, and speeding up the training and inference process.[55] Additionally, recent work has demonstrated that integrating equivariant layers into a transformer framework can further improve predictive power.[56–58] These architectures have achieved state-of-the-art performance across diverse domains requiring geometric fidelity, including molecular dynamics[56,57,59–61], charge density prediction[62,63], accurate formation energy prediction[64], and protein structure prediction[65]. The same geometric deep learning principle should naturally extend to the modeling of reticular materials, in which the strength of equivariant networks remains unexplored.

Beyond predictive accuracy, interpretability can be valuable in materials modeling because it may help relate model predictions to chemically meaningful structural features and generate testable hypotheses about structure-property relationships.[66–68] Being a topic still in infancy, explainable machine learning or artificial intelligence (XML or XAI) has drawn notable attention from the materials science modeling field, with several approaches being introduced. Feature importance methods, such as SHAP, can help identify the most influential features in predictions.[69,70] However, these importance numbers can be challenging to interpret when features are correlated or when dealing with complicated problems. Another way to increase model explainability is to select some typical examples from the input, using methods like k-nearest neighbors.[66] Intuitiveness can be provided by showing similar training samples, but the choice of examples can limit their strength and may not generalize well. Surrogate models approximate complex models with simpler, more explainable ones, offering a route to interpretability.

However, this often comes at the cost of predictive performance, and the choice of the surrogate model is frequently subjective.[71–73] Parameter or intermediate output inspection techniques (*e.g.*, t-SNE and PCA) allow researchers to visualize the model's learned embeddings (*e.g.*, SchNet), offering insights into the data's structure. Still, these visualizations are usually subjective and may not directly lead to actionable understandings.[74–77] Therefore, given the generally limited clarity of materials data, building an ML model for materials design with well-balanced predictive power and interpretability is still under investigation.

In this work, we introduce a site-resolved equivariant learning framework for reticular materials. We propose a three-dimensional periodic-space sampling method that decomposes periodic structures into local geometric environments, enabling simultaneous property prediction and site-level contribution learning from structure-level supervision. This formulation allows the model to identify physically meaningful local environments that govern material properties and to quantify their contributions in a unified framework. To train and test our model, we retrieve all previous databases and build a new binary adsorption dataset on $CO_2$ and $N_2$ in CoRE MOFs.[78–80] Our model outperforms state-of-the-art ML models with superior predictive accuracy and data efficacy for exemplified applications (*i.e.*, gas storage, separation, and electrical conduction). Moreover, we demonstrate that our model understands the local geometrical features and clearly establishes the connection between structure and properties. The corresponding strong sites for each targeted application are summarized and further translated into a motif-guided forward-design workflow, in which high-contribution local motifs are assembled into new reticular frameworks and validated by physics-based calculations. This proof-of-concept yields representative newly designed MOFs with record-high performance across gas storage, gas separation, and electrical conduction. The resulting framework provides a unified approach to accurate prediction, site-resolved interpretability, and data-driven materials design and is released as an open-source tool for broader application to reticular and crystalline materials systems.

**Equivariant graph neural networks for interpretable reticular framework design**

The performances of reticular frameworks are dictated by their diverse and structurally complex architectures. Just as adsorbate molecules do not distribute uniformly throughout framework pores, different regions or local sites within these frameworks can contribute very differently to overall performance. For example, open metal sites (OMSs) in MOFs facilitate strong interactions with various molecules and dominate total gas adsorption[81]. Tetrathiafulvalene sites in COFs promote the formation of π-conjugated systems, enabling high electrical conduction[82]. Therefore, identifying the strong performance sites in reticular frameworks is critically essential to obtain the capability to fine-tune those sites, sterically and electronically, for maximized performance output. However, experimental determination of the strong sites is rather challenging due to the limitations of current characterizations in providing adequate structural details of the sites[83,84]. Computational methods such as grand canonical Monte Carlo (GCMC) and density functional theory (DFT) generally demand a relatively large amount of computational resources to obtain[80,85]. Here, we introduce a site-resolved learning framework that decomposes periodic structures into local geometric environments and learns their contributions to global properties in a weakly supervised manner. Additionally, it partitions the structure to reduce input dimensionality, thereby enhancing model accuracy and transferability in low-data regimes. By benchmarking the contributions of distinct sites, we can identify superior sites, analyze the factors that make them exceptional, and help determine how to combine them to design advanced reticular materials.

Specifically, we combine a periodic local-environment sampling scheme with an established E(3)-equivariant transformer-style architecture to learn site-level contributions from only structure-level labels in an inexactly supervised manner. This design allows the model to preserve 3-D geometric information while producing explainable site-wise attributions across periodic reticular frameworks. As shown in **Fig. 1a-d**, we sample the periodic structure in uniform density by centering a spherical detection window of radius $r$ at a set of Cartesian sampling points $c_i$ distributed through the unit cell with the same real-space step size $d$. Each window defines a local site $S_i = \{(Z_j, \mathbf{r}_j) \mid \| \mathbf{r}_j - c_i \|_{\mathrm{PBC}} < r\}$, where $Z_j$ and $r_j$ denote the atomic species and Cartesian coordinate of the atom $j$, respectively, and the distance is evaluated under periodic boundary conditions. Each sampling point serves only as an anchor for a variable-sized local atomic environment, which is then processed independently by the equivariant model. At each training epoch, sampling is performed in Cartesian space beginning at a random position inside the cell, so that it can efficiently exploit the structure even for non-orthogonal unit cells.

As shown in **Fig. 1e**, for each site $S_i$, the model input is $x_i = \{(Z_j, \mathbf{r}_j)\}_{j \in S_i}$, namely, the set of atomic species and Cartesian coordinates inside the detection window. There are no hand-crafted descriptors or precomputed energy fields supplied. These local point-cloud inputs are embedded through decomposing into radial features and irreducible angular geometric representations[57] within the E(3)-equivariant transformer backbone, which inherently enforces rotational, translational, and inversional invariance for robust, physically consistent feature extraction. Conceptually, the equivariant transformer follows the format of a standard transformer encoder (detailed architecture in **Fig. S1** and **Algorithm S1**), with radial and angular embedding layers[55,58] to lift the inputs to the latent space of irreducible representation, then stacked self-attention blocks[57,86] capturing the pairwise and higher-order interactions, following a MLP contribution layer to predict the scalar contribution for a specific site. The network predicts a scalar contribution $y_i$ for each sampled site. The structure-level prediction is then obtained by averaging over all N sampled sites: $\hat{y} = (1/N)\Sigma_{i=1}^{N} y_i$.

**Datasets construction and model training & optimization**

We demonstrate our model for three exemplified applications of reticular frameworks (MOFs): $N_2$ adsorption, $CO_2/N_2$ separation, and electrical conduction. For $N_2$ adsorption in MOFs, the adsorption capacity data in the CoRE MOF 2019 database[88] were retrieved with properties at 1 bar pressure and 77 K in the isothermal results.[78–80,88] Reticular frameworks are promising for gas separation applications due to their unique properties, such as well-defined pore structures and tunable surface chemistry, enabling precise molecular sieving and selective adsorption[89,90]. However, ML applications for gas separation prediction in reticular frameworks were constrained by the lack of high-quality datasets and the general difficulty of mixed gas adsorption computation[26,34]. Accurate prediction of mixed gas adsorption demands that models genuinely capture how each local site in the framework interacts with different gas species, rather than relying on simplified proxies like surface area. Here, we construct our dataset by simulating $CO_2/N_2$ binary gas adsorption in CoRE MOF 2019 structures[79] at a temperature of 298 K, total pressure of 1 bar, and the ratio of $CO_2$ to $N_2$ fixed at 0.15:0.85[22], using gRASPA,[91] a graphics processing unit (GPU)-accelerated Monte Carlo software. This new consistent dataset enabled us to evaluate our model performance by discerning the local sites' contributions to the overall separation of two gases.[91] Finally, to assess the model's versatility, we evaluated its capability on bandgap prediction using the retrieved QMOF dataset[80]. Distributions of the four datasets are illustrated in **Fig. S2**.

For each dataset, materials were partitioned into training, validation, and held-out test sets in 70/15/15 division using a fixed random seed. Hyperparameters were selected using only the training and validation sets, and all final benchmark metrics reported in the main text are evaluated on the held-out test set. Among all hyperparameters, detection window (site) radius and sampling step size are critical for capturing the local chemical environment. The window radius must be appropriately set to balance the need for sufficient local chemical information against the risk of unnecessarily increased input dimensionality. An undersized window may fail to encompass the relevant pore and channel accessibility information, while an oversized window could dilute the local features critical to adsorption. To address this, we conduct a convergence test with different window radii (including the whole structure) as input for targeted applications (see **Fig. S3**). Our results indicate that a radius $r$ of 6 Å/8 Å is optimal for gas adsorption and band gap prediction. Similarly, the sampling step size is set to $r - 2$Å for the trade-off between accuracy and computational cost. Furthermore, to ensure a fair comparison when benchmarking our approach against state-of-the-art ML algorithms for reticular frameworks, we use a fixed random seed when partitioning the datasets for all the models, and thoroughly optimize hyperparameters of these reference models with current datasets employing the Optuna library.[92]

**Benchmarking against state-of-the-art models and data efficacy**

As shown in **Tab. 1** and **Fig. 2**, under the three-run random-split evaluation protocol with early stopping after 20 epochs, our model X(3)mat demonstrates outstanding accuracy and robust data efficacy against state-of-the-art models for reticular frameworks across all applications (*i.e.*, $N_2$ adsorption, $CO_2/N_2$ separation, and electrical conduction). For MOF $N_2$ adsorption prediction, our model outperforms all other models with a large gap (**Fig. 2a,d**), which is fed with extra energetic information from additional calculations. For $CO_2/N_2$ separation and band gap predictions in MOFs (**Fig. 2b,c,e-g**), our model once again exhibits superior precision against existing models. Furthermore, we evaluate the data efficacy of our model against current leading methods (*e.g.*, MOFormer[31] and CGCNN[31]). As shown in **Fig. 2h**, our model demonstrates improved data efficiency with dataset sizes from $10^2$ to $10^4$. We attribute its performance to equivariant networks' capability to reduce the redundant freedoms while preserving the key symmetries, and the usage of the local structure within a detection window as input. Such a dimensionality reduction is particularly advantageous when working with limited sample sizes.

**Demonstrating model interpretability in exemplified applications**

*Gas adsorption and separation*

The most prominent feature of our method, in contrast to previous methods, is its ability to detect the local geometries in the structures that contribute to the target property rather than solely predict the overall property value. The output of our model provides a smooth map for illustrating the contributions of all local sites within a reticular structure. As shown in **Fig. 3a**, overlaying the spatial map of site-wise contributions ($y_i$) as predicted by our model with the GCMC-computed $N_2$ adsorption density (Y) distribution reveals a pronounced correlation. Using our model, we can assign the adsorption sites ($site_i$) surrounding these high-density zones that are most responsible for the observed $N_2$ uptake. The 4,4′-bipyridine of the large pores in MOF-KUZPEC are flagged by our model as the dominant adsorption sites ($site_i$) with maximized contribution ($y_i$ = 306.34 g/L). The joint effect of two nearby strong adsorption sites brings forward the adsorption density maximum as observed (grey arrows). More examples are given for more MOFs in **Fig. S5.**

For gas separation/mixed-gas adsorption, each site's contributions to $CO_2$ and $N_2$ adsorptions are predicted by our model, respectively. Then, the spatial selectivity contribution map can be computed by taking the ratio of these contributions. As shown in **Fig. 3b**, three distinct strong adsorption sites ($site_i$, $site_j$, $site_k$) are identified for both $N_2$ adsorption ($y_i/y_j/y_k$ = 2.42/2.30/1.23 g/L) and $CO_2$ adsorption ($y_i/y_j/y_k$ = 396.15/414.10/477.12 g/L). The contribution of $N_2$ is confined to regions distant from the site due to its relatively large molecular size. By contrast, $CO_2$ exhibits strong contributions distributed across the site, with maxima near the Gd metal nodes. The selectivity maximum of $CO_2$ over $N_2$ is then located in the regions with strong $CO_2$ contribution yet relatively weak $N_2$ contribution. The whole procedure is visually analogous to "carving out" the strong $N_2$ contribution regions (black arrows) from the $CO_2$ contribution map (yellow arrows), and the resulting selectivity contribution distribution closely mirrors the ground truth selectivity pattern (orange arrows). The capability of our model in determining the highly selective sites applies universally across the dataset, and more examples are given in **Figs. S6** and **S7**.

*Electrical conduction*

Conductive reticular frameworks (*e.g.*, MOFs and COFs) are intriguing due to their potential to combine the conventional reticular materials' merits, such as porosity and tunable chemistry, with electrical conductivity, which makes them promising for a wide range of applications, including energy storage, catalysis, and sensing. Many approaches have been put forward to realize electrical conductivity in reticular materials, including networks of coordination bonds for continuous charge delocalization [93,94], extended π–d conjugation crossing the planes[95,96], stacked π–π interaction through the space[97], redox hopping[98], and so on[80,99,100]. However, there is a lack of quantitative tools for the accurate identification of conductive channels/sites and evaluation of their contributions to the overall electrical conductivity, where our model can fill the gap. As shown in **Fig. 4a,b**, deeper blue colors indicate lower band gap contributions, or in other words, higher electrical conduction contribution, in the contribution distribution maps predicted by our model. For the framework with a high band gap (5.11 eV) and low conductivity (**Fig. 4a**), its strong conduction contribution sites barely appear on the acylammonium functional group with a relatively low band gap contribution ($y_i$ = 4.57), corresponding to the slightly delocalized yet isolated electron density around the acylammonium functional group (grey arrow and **Figs. S8a** and **S9a,b**).

In contrast, for the MOF with a zero band gap (**Fig. 4b**), the strong conduction sites of the cobalt node exhibit a negative band gap contribution ($y_i = -0.40$), corresponding to the well-delocalized electron density around the bridging enediolate cobalt motif (black arrow). What is more important is that all these strong conduction sites are bridged together in the reticular framework (**Figs. S8b** and **S9c,d**), indicating a continuous electron transport pathway and then a metallic band gap. Another example of using our model for conductive MOF analysis is shown in **Fig. S10**. Although band gaps in periodic solids involve delocalized electronic states, band-gap variations in MOFs are often strongly governed by local-to-medium-range structural motifs, including metal-node electronic structure, metal-linker energy matching, orbital overlap, linker conjugation, and through-bond connectivity. The detection window used for band-gap prediction is sufficiently large to capture multiple coordination shells and extended node-linker environments, which encode key features controlling orbital energies and effective electronic coupling. Thus, while X(3)mat does not explicitly decompose fully delocalized Bloch states, its local-environment representation captures the dominant structural factors that govern band-gap trends in MOFs.

Delving further, our model can generate a histogram showing the distribution of local sites with strong, medium, and weak contributions, which will also elucidate the structural origins of their electronic properties. As depicted in **Fig. 4c**, charge conduction in this example structure is mainly from the Co node and the connecting cyclobutanetetrone bridge, which form the charge transport pathway of the structure. Meanwhile, many high-frequency local sites react against the conduction with positive bandgap contributions. Such a contribution distribution histogram can also be extended for the entire dataset, with an extra axis indicating the real band gap range of all MOF structures. In **Fig. 4d**, a notable peak ridge shows up (along $y = x$), signifying that for all structures sharing a given band gap, their ensemble of site-wise contributions is distributed with its peak precisely at the actual band gap value. Additionally, the region of negative contributions shows the strong sites that act to lower the overall band gap, and these sites are of interest for future conductive MOF design. The area of this region decreases as the real band gap increases. Finally, it disappears when the real band gap approaches around 4.5 eV, indicating the vanishing of local conductive sites in highly dielectric MOFs. The same histograms depicting $N_2$ contribution distribution in MOFs, $CO_2/N_2$ separation in MOFs, are shown in **Figs. S11**, **S12**, and **S13**, respectively.

**Identification of prevalent strong contribution sites and design principles**

The rational design of reticular frameworks targeted at specific properties by tuning their local geometries/sites is crucial for various applications. It is then compelling to survey all the existing local geometries and pinpoint the ones that contribute the most to specific applications. Reticular frameworks with improved performance can then be developed, taking advantage of these top candidates. To achieve this goal, we should be able to decompose and distinguish various types of local sites before quantifying their contribution and ranking them subsequently. Our model has demonstrated its capability in decomposing a reticular structure and evaluating the contributions of the resulting components. How to determine if two sites are identical/distinct, given that all atoms at these sites are in constant vibration, is the problem to solve here. To address this, we eliminate the coordination variables of the atoms and use the graph structure as the identity of the site. For each pair of sites, we compute their Weisfeiler-Lehman (WL) graph hash values[101]. If the hash values match, they are considered the same graph and therefore the same site. For reticular frameworks, H-removed graphs are used to save computation resources and reduce the complexity of the graph representation while preserving the site's topological information[102].

**Tab. S1** presents the strong adsorption sites identified in the CoRE MOF dataset. Site 1 and Site 2 correspond to an open metal site (OMS) of Zr and Cu surrounded by benzene or pyridine rings. Site 3 is two benzene rings oriented perpendicular to each other near one OMS of Cu. Site 4 is an OMS of Zn forming a plate. The plate-like local structures with hollow regions perpendicular to them can serve as decent landing sites for the adsorbate, and therefore significantly increase the adsorption contributions. Site 5 is an OMS of Mg connected by a benzene ring stacked over another benzene ring. All these top sites confirm the dominance of OMS upon $N_2$ adsorption, while sites 3 and 5 highlight the significant role of benzene rings in nitrogen adsorption under 1 bar pressure[103].

The top 5 $CO_2$-selective sites are shown in **Tab. S2**. Site 1 is an OMS of Zr metal sites connected with multiple benzene ring linkers. Similarly, site 5 in Mg-MOF-74 (CCDC code: VOGTIV) consists of two groups of Mg metal sites connected together, which has been reported by Britt et al. as a well-known $CO_2$ strongly selective capture site owing to the intrinsic binding nature of the open metal centers[104]. Site 2 also corresponds to a $CO_2$ selective adsorption site in a representative reported structure[105], attributable to the presence of abundant amino groups. These $–NH_2$ moieties, together with their appended amine functionalities, significantly enhance $CO_2$ selectivity by functioning as Lewis base centers and providing the hydrogen bonding to oxygen atoms of $CO_2$. Site 3 corresponds to the Co metal center, from which the linkers bearing phenyl rings extend outward in four directions. Site 4 is a pore enclosed by C–N heterocycle linkers and Eu metal nodes. This pore of limited size selectively excludes the $N_2$ molecules, yet accommodates the adsorption of the smaller $CO_2$ molecules[106]. In addition, the attraction exerted by the Eu nodes further enhances the high $CO_2$ selectivity observed at this site.

The top five conductive sites as identified are shown in **Tab. S3**. Sites 1 form conductive pathways through the stacking heterocyclic rings composed of carbon and nitrogen atoms. It highlights that nitrogen-substituted aromatic linkers, such as triazine, offer intrinsic advantages over simple benzene rings by promoting extended π-conjugation, enforcing geometric planarity[107], and enhancing π-acidity[99,108], thereby strengthening charge transport *via* both π–d conjugation and π–π stacking pathways. Meanwhile, sites 2, 3, and 5 highlight the advantage of M–N coordination (pyrazolate/iminyl/diamine donors) over M–O (carboxylate/phenolate) for electronic transport in MOFs. Compared to the relatively hard O donors, softer N donors afford greater metal–ligand covalency and better energy matching with aromatic π-orbitals, yielding dispersive bands and higher carrier densities[100]. Sites 4 and 6 are large, coplanar fused aromatics. They maximize both the through-bond conjugation across the metal–linker lattice and the through-space π–π overlap between stacked sheets, thereby broadening bands and shrinking transport gaps[109,110]. All these top contribution sites can then be employed for the design of advanced reticular materials targeted at diversified applications.

### X(3)mat-guided motif transfer and forward design of MOFs

To translate the learned strong-site motifs into new materials (workflow shown in **Fig. S15**), we first constructed a ranked library of high-contribution motifs identified by X(3)mat, and derived the corresponding nodes and linkers either by extracting them from parent frameworks. These building units were then assembled across candidate topologies, filtered according to whether the resulting pore sizes could reproduce local environments similar to the original strong sites, and screened again by X(3)mat before physics-based validation. A representative $N_2$-storage design example is shown in **Fig. 5a**, where two high-contribution local motifs are combined into a new reticular framework. After geometry optimization and GCMC validation, all $N_2$-storage designs reported here exhibit uptakes higher than the maximum value in the CoRE MOF 2019 dataset under the same simulation conditions, demonstrating that the learned strong sites can guide the construction of frameworks beyond the performance range of the data used for model development (**Fig. 5b**). Applying the same design principle to other targets, we further obtained newly designed MOFs with strong $CO_2/N_2$ separation performance and sub-0.2 eV band gaps, the representative candidate details provided in **Fig. 5c.** Together, these results show that X(3)mat-derived strong-site motifs can serve as transferable design elements for property-oriented reticular frameworks discovery.

### Conclusions

We introduce X(3)mat, an E(3)-equivariant graph learning framework for interpretable representation learning in reticular materials. By decomposing periodic structures into local geometric environments through a three-dimensional periodic-space sampling strategy, the model enables unified learning of global material properties and site-resolved contributions within a single architecture. Trained on existing datasets and a newly constructed CoRE MOF $CO_2/N_2$ separation dataset, X(3)mat achieves state-of-the-art accuracy and data efficiency for gas storage, gas separation, and electronic-property prediction. The learned high-contribution motifs were further translated into 10 newly designed MOFs with ultrahigh $N_2$ storage, strong $CO_2/N_2$ separation performance, and sub-0.2 eV band gaps, all independently validated by GCMC or DFT calculations. More broadly, because the representation operates directly on periodic atomic environments, X(3)mat provides a general and extensible approach for interpretable and design-oriented machine learning across reticular and crystalline materials systems.

## Methods

*1. Hyperparameter optimization of the reference models*

Exact model architectures of the prior studies are retrieved from the original papers. MLP models were employed to reproduce the previous persistent homology, 2-D energy histogram, and 1-D energy histogram approaches' results. The energy bin width and energy range are fully optimized for producing the 1-D and 2-D energy histograms. The gradient bin width and gradient bin width are 25 kJ/mol/Å and 0~150 kJ/mol/Å, respectively. Layer numbers of the MLP are fixed to 4 for the persistent homology, 1-D, and 2-D models, following previous studies[34,37]. To reproduce the 3-D voxel method results, the 3-D convolution ResNet is employed as in the original study[40]. Hyperparameters of these models and the CGCNN model[31] are fully optimized *via* the TPESampler algorithm implemented by optuna[92,111] . Notably, we did not re-tune MOFormer's hyperparameters other than the fine-tuning learning rate, so that the MOFormer could be initialized from the self-supervised pretrained checkpoint provided by the authors[31], and the learning rate was selected from $\{1 \times 10^{-5}, 5 \times 10^{-5}, 1 \times 10^{-4}\}$ using the validation set. Complete hyperparameter tables are now provided for X(3)mat and all baseline models in **Tab. S7 and S8** in the Supporting Information, including the search ranges and optimized values of the architecture-level parameters (*e.g.*, number of layers, hidden dimensions, attention heads) and training settings (*e.g.*, learning rate, batch size).

*2. Open dataset construction for MOF $CO_2/N_2$ separation*

The adsorption capacity of $CO_2/N_2$ binary mixture in MOFs at 298 K, 1 bar, and a molar ratio of 0.15:0.85, was calculated using grand canonical Monte Carlo (GCMC) simulations using the gRASPA software (version 2025-Mar09).[91] Crystal structures of CoRE MOF 2019 were downloaded from the MOFX-DB database.[88,112] Structures containing extra-framework ions were removed from the dataset. To implement a generalized workflow that retrieves the crystal structures from a database, we developed a Python package, namely gRASPA_job_tracker (version 2025-06-23). The details of this package are discussed in the **Supplementary Notes**. During the simulation, the framework was kept rigid and non-bonded interactions were modeled using the Lennard-Jones (LJ) potential with a 12.8 Å cutoff with cross-interaction terms estimated using Lorentz–Berthelot combining rules. Universal force field (UFF) parameters were applied to MOF atoms, while $CO_2$ and $N_2$ were treated using the TraPPE force field.[113–115] Long-range electrostatics were computed *via* Ewald summation with a precision of $1\times10^{-6}$.[116,117] Partial atomic charges for MOFs were predicted by the PACMOF2-neutral machine learning model (DDEC6-based), and those for adsorbates were taken from TraPPE.[118,119] [114] Complete information about the simulation can be found in the **Supplementary Notes A**. Notably, we aim to ensure reproducibility by defining all the required simulation parameters and loading dependencies through a centralized configuration file, which is discussed in **Supplementary Notes** and must be available with any dataset generated from gRASPA_job_tracker.

*3. Reticular framework electronic structure calculations from first principles*

Density functional theory (DFT) calculations were conducted using Vienna Ab initio Simulation Package (VASP)[120] version 5.4.4. The Perdew-Burke-Ernzerhof (PBE)[121] exchange-correlation functional within the Generalized Gradient Approximation (GGA)[121] was employed for its computational efficiency and reliability. Projector-augmented wave (PAW) pseudopotentials, as recommended for VASP v.5.4.4, were used for all elements[122,123]. A plane-wave energy cutoff of 550 eV was applied, and all calculations were spin-polarized. We used VASPKIT[124] to set up the calculations with the Monkhorst-Pack k-point grids and a density of 0.03 to ensure adequate sampling of the Brillouin zone.

**Data availability**

Data for the training of the X(3)mat, including the dataset, GCMC movies, and the details of $N_2$ GCMC simulation input files, are available at Zenodo with DOI: 10.5281/zenodo.20050356.

GCMC simulation input files for $CO_2$/$N_2$ mixture adsorption in MOFs are provided in https://github.com/Shi-Research-Group/gRASPA_job_tracker/tree/main/examples/data/coremof_clean.

**Code availability**

Code for the X(3)mat is available at https://github.com/zhenpengyao/X3MAT.

**Acknowledgements**

Z.Y. and Z.Z. were supported by the National Natural Science Foundation of China (grant number 52373228). Z.Y. and K.S. thank Prof. Andrew Rosen for helpful discussions on MOF electrical conduction. K.S. thanks Dr. Zhao Li for beneficial discussions on the gRASPA software. K.S. acknowledges the partial support from the ACS Petroleum Research Fund under Doctoral New Investigator Grant 68604-DNI6. This work used the Delta system at the National Center for Supercomputing Applications through allocation CHM240039 from the Advanced Cyberinfrastructure Coordination Ecosystem: Services & Support (ACCESS) program, which is supported by National Science Foundation grants #2138259, #2138286, #2138307, #2137603, and #2138296.

**Contributions**

Z.Y. conceived the overall project and supervised the research. Z.Z. built the model architecture, performed the $N_2$ Adsorption simulations, and conducted the data analysis. S.B.K. and K.S. prepared the new MOF $CO_2$/$N_2$ separation dataset. Z.Z., Z.Y., and K.S. contributed to the development of model interpretability. All authors contributed to the preparation of the manuscript.

**Competing interests**

The authors declare no competing interests.

# Figures

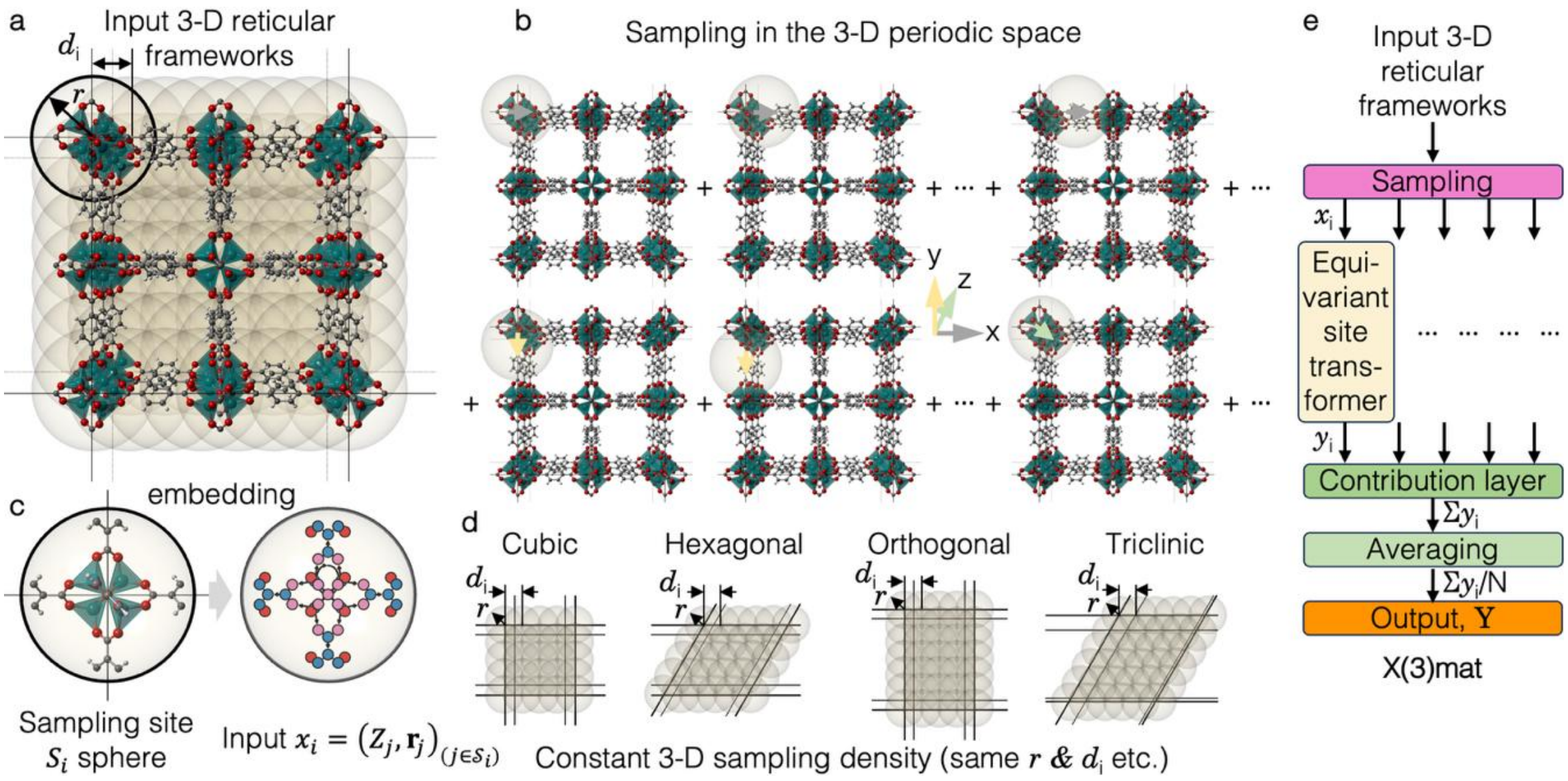

**Figure 1. Equivariant graph neural networks for interpretable design of reticular frameworks. a.** Arrangement of the detection windows in the exemplified input structure (*e.g.*, MOF-5) in spacing $d$ . **b.** Illustration of the sampling process throughout the input structure with radius $r$. **c.** An example of the sampling site and its encoded 3-D graph inside a detection window. **d.** The behavior of the sampling process in distinct crystal systems. **e.** Structure of the equivariant graph neural networks X(3)mat for interpretable design of reticular frameworks. $x_i$ denotes the local atomic environment extracted at the sampling point $i$, $y_i$ denotes the corresponding predicted site contribution, and the structure-level prediction is obtained as the average over $N$ sampled sites.

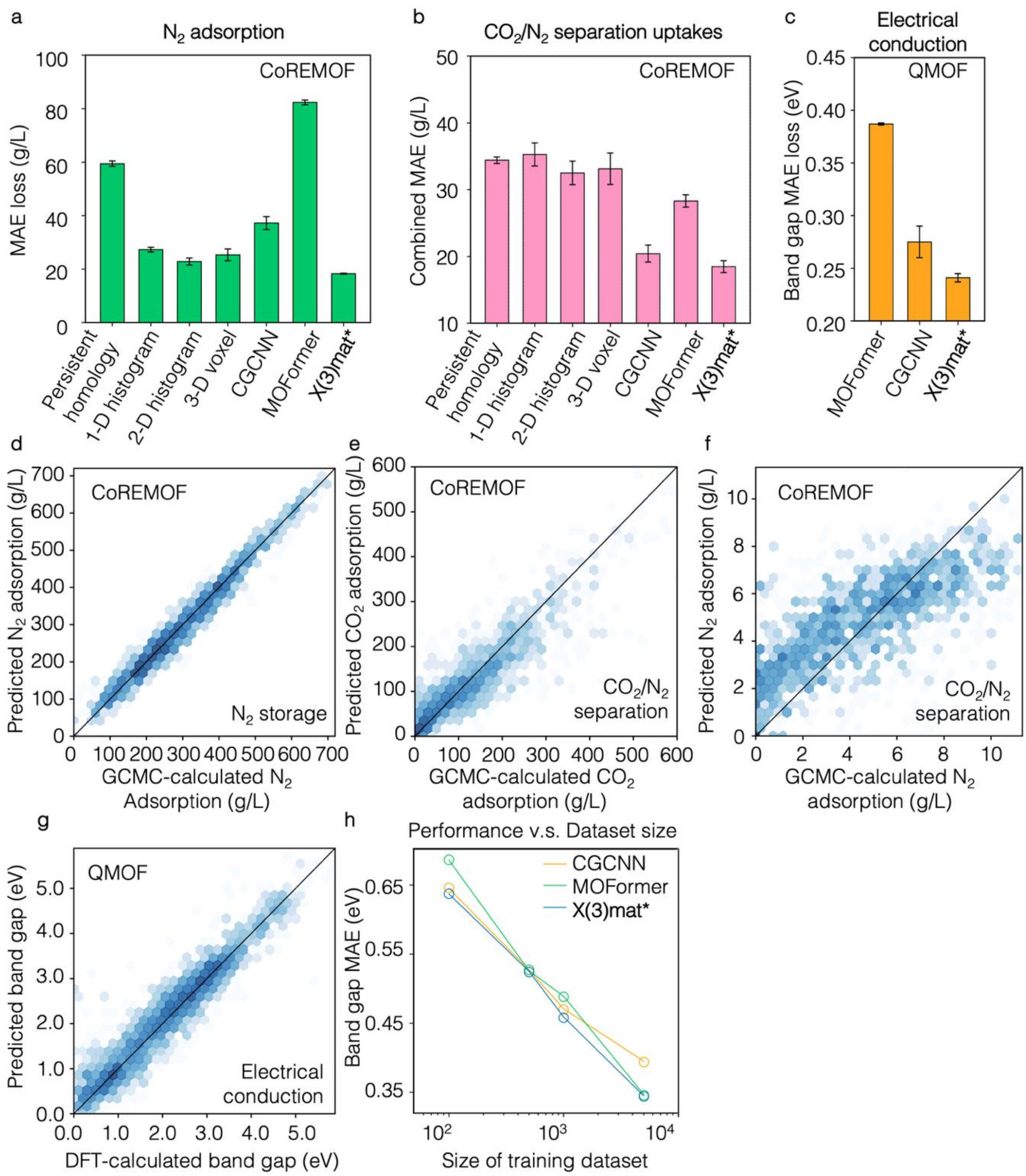

**Figure 2. X(3)mat enables efficient and accurate prediction of reticular frameworks' properties in exemplified applications. a-c**, Benchmark performance for $N_2$ adsorption in CoRE MOF 2019, $CO_2/N_2$ separation in CoRE MOF 2019, and band-gap prediction in QMOF, respectively. **d-g**, Parity plots comparing X(3)mat predictions with GCMC or DFT reference values for $N_2$ adsorption, $N_2$ and $CO_2$ uptake under $CO_2/N_2$ separation conditions, and band gaps. **h**, Data-efficiency benchmark on the QMOF dataset.

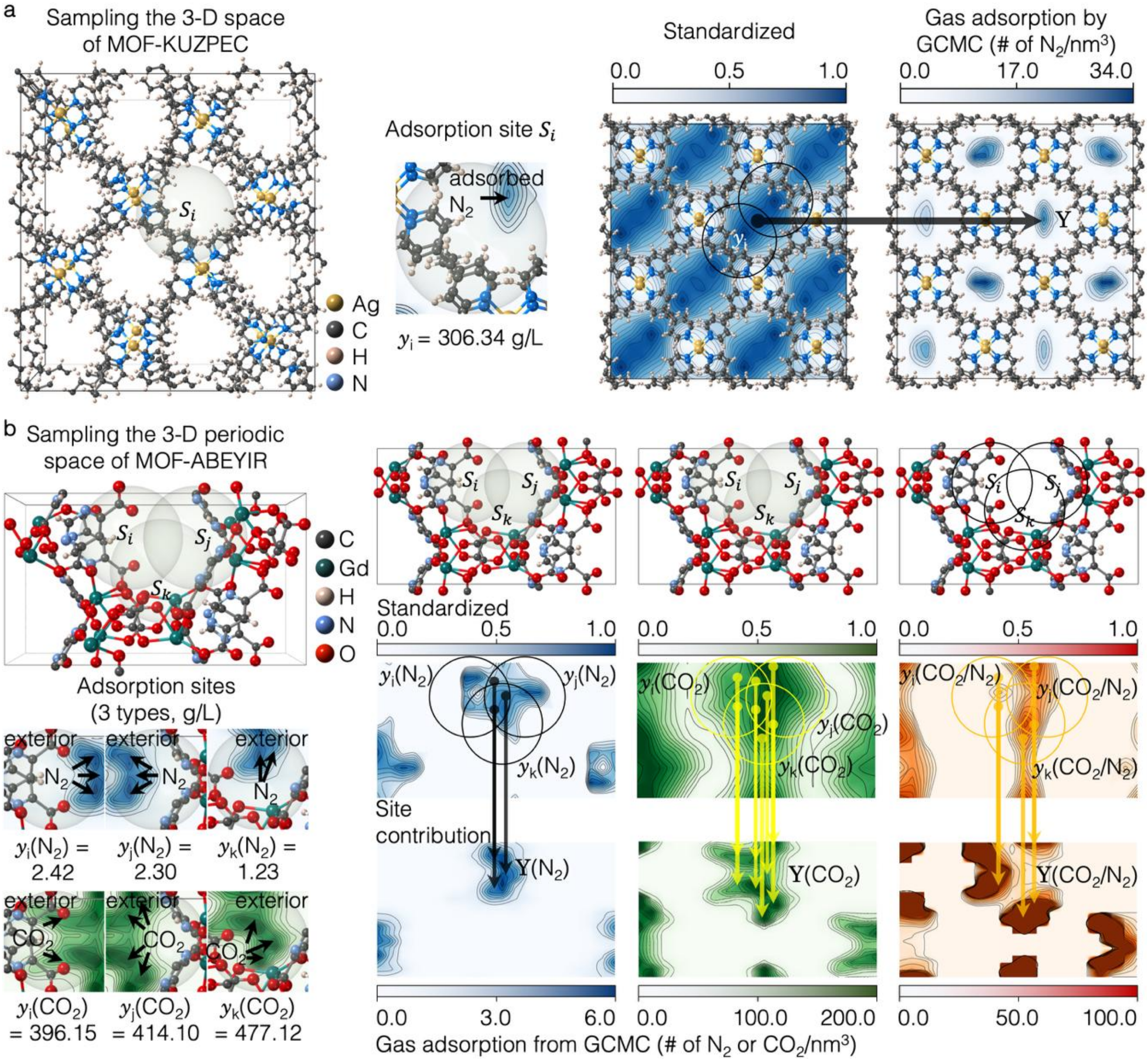

**Figure 3. X(3)mat interpretability demonstration in reticular frameworks design for gas adsorption and separation. a.** Spatial map of site-wise contributions predicted by X(3)mat in MOF-KUZPEC shows pronounced correlation with the GCMC computed $N_2$ adsorption density distribution. The model flags cavity wall edges as the dominant adsorption sites (site$_i$) with maximized contribution, and the joint effect of two nearby strong sites brings the adsorption density maximum as observed (black arrows). **b.** For gas separation, each site's contributions to $CO_2$ and $N_2$ adsorptions are predicted by our model, and the spatial selectivity contribution is then computed through ratio calculation. Three distinct strong adsorption sites (site$_i$, site$_j$, site$_k$) are identified for both $N_2$ and $CO_2$ adsorption. The selectivity maximum of $CO_2$ over $N_2$ is then located in the regions with strong $CO_2$ contribution yet relatively weak $N_2$ contribution. The whole procedure is visually analogous to "carving out" the strong $N_2$ contribution regions (black arrows) from the $CO_2$ contribution map (yellow arrows), and the resulting selectivity contribution distribution closely mirrors the ground truth selectivity pattern (orange arrows).

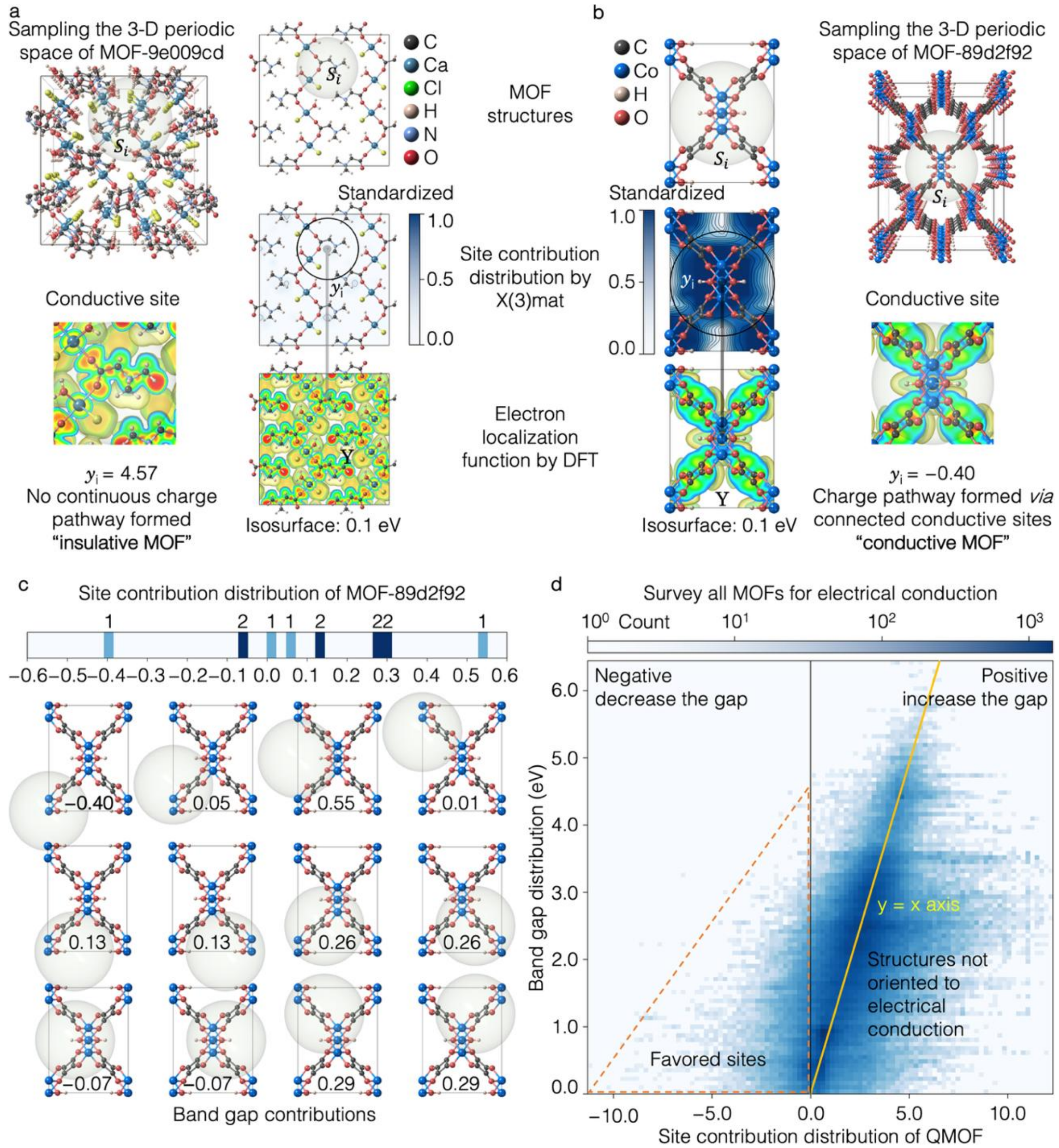

**Figure 4. X(3)mat interpretability demonstration in electrically conductive MOF design. a**, Comparison between DFT isosurface maps and site-resolved band-gap contribution densities (higher transparency indicates a larger band gap). For a wide-band-gap MOF (5.11 eV), low-contribution sites are weakly localized on the acylammonium group ($y_i$ = 4.57), consistent with localized electron density (grey arrow). **b**, Same comparison for a zero-band-gap MOF, where high-contribution conductive sites on the Co node exhibit negative band-gap contributions ($y_i$ = −0.40), consistent with delocalized electron density around the bridging enediolate–Co motif (black arrow), forming a continuous transport pathway. **c**, Distribution of site contributions across strong, medium and weak conductive environments, highlighting structure–property origins of electronic behavior. **d**, Correlation between site contributions and band gaps across the QMOF dataset. The diagonal ridge ($y = x$) indicates alignment between peak site contribution and global band gap, while negative contributions identify motifs that reduce band gaps and enable conductivity design.

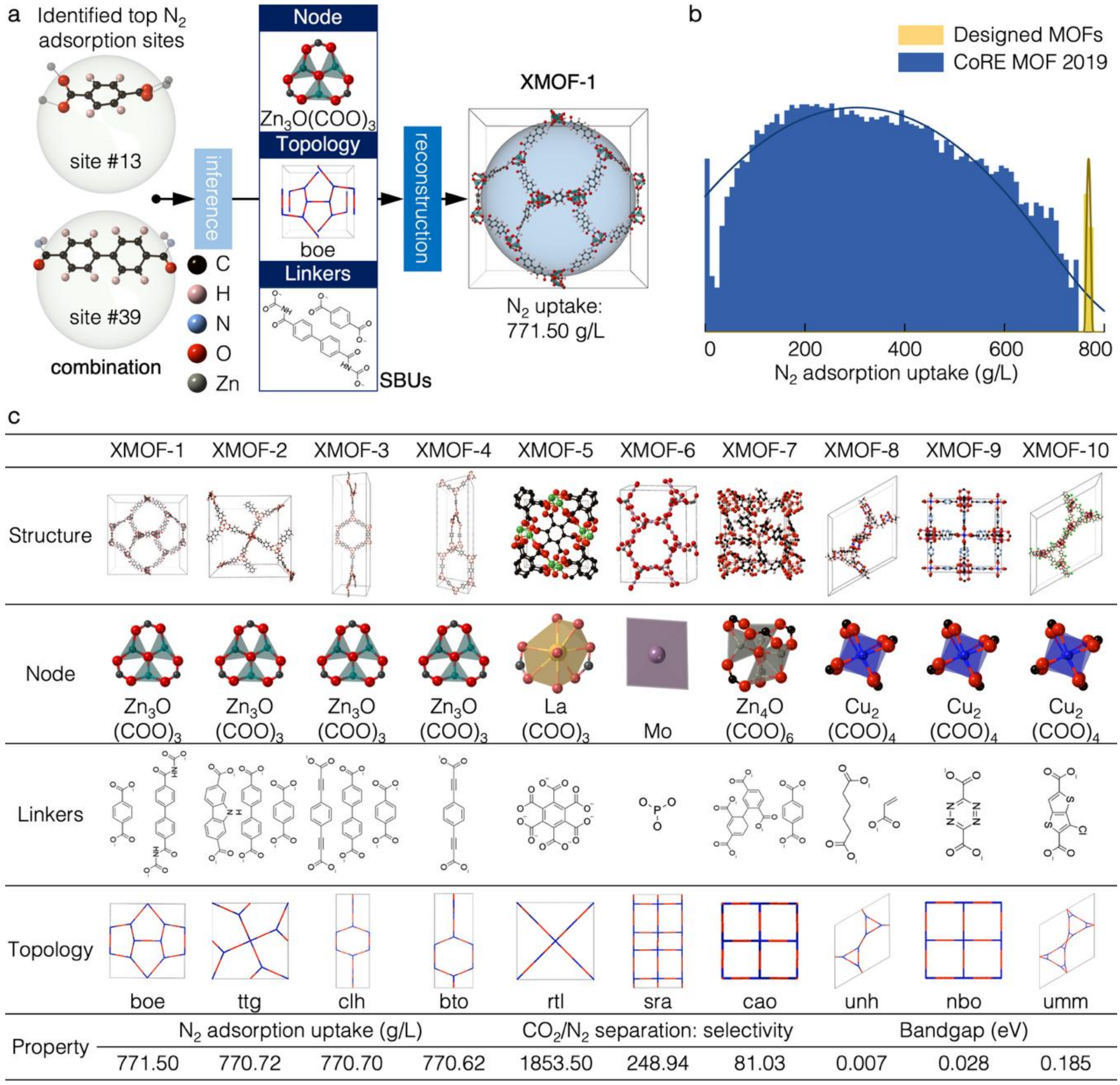

**Figure 5. X(3)mat-guided motif transfer and forward design of MOFs for gas storage, gas separation, and electronic properties. a.** Representative motif-guided forward-design example for $N_2$ storage. Two high-contribution adsorption motifs identified by X(3)mat are selected from the strong-site library and assembled into a chemically compatible reticular framework, yielding a newly designed MOF candidate. The panel illustrates that the design step is guided by transferable local motifs rather than by direct copying of a parent structure. **b.** Distribution of GCMC-validated $N_2$ uptakes for newly designed MOFs generated by the motif-guided workflow. All candidates were proposed from X(3)mat-identified strong adsorption motifs, screened by the model, optimized by VASP, and subsequently validated by RASPA2 GCMC at 77 K and 1 bar. The vertical dashed line indicates the highest $N_2$ uptake observed among the structures in the CoRE MOF dataset, and the designed candidates lying to the right of this line therefore exceed the best uptake among the known data of the model. **c.** Summary of representative validated XMOF candidates generated by the motif-guided workflow. The table lists the optimized structures, nodes, linkers, topologies, and target properties of ten designed MOFs.

# Tables

**Table 1. Benchmarking against state-of-the-art reticular framework models.** All baseline results were reproduced in this work using the cited models but optimized hyperparameters for the consistent data in this work, with the exception of MOFormer and CGCNN on the QMOF band-gap task, for which the original implementations were directly adopted.

| | MOF $N_2$ storage | | MOF $CO_2/N_2$ average uptake | | MOF band gap | |
|---|---|---|---|---|---|---|
| | MAE (g/L) | $R^2$ | MAE (g/L) | $R^2$ | MAE (eV) | $R^2$ |
| Persistent homology[34] | 59.44±1.03 | 0.663±0.004 | 34.42±0.48 | 0.555±0.026 | - | - |
| 1-D energy histogram[35] | 27.28±0.91 | 0.932±0.002 | 35.27±1.74 | 0.587±0.018 | - | - |
| 2-D energy histogram[36] | 22.81±1.31 | 0.952±0.005 | 32.50±1.78 | 0.631±0.022 | - | - |
| 3-D voxel[40] | 25.33±2.21 | 0.618±0.483 | 33.12±2.36 | 0.266±0.248 | - | - |
| CGCNN[31] | 37.23±2.43 | 0.774±0.017 | 20.42±1.27 | 0.801±0.024 | 0.275±0.015 | - |
| MOFormer[31] | 82.36±0.89 | 0.353±0.006 | 28.30±0.92 | 0.682±0.009 | 0.387±0.001 | - |
| **X(3)mat*** | **18.30±0.17** | **0.969±0.001** | **18.46±0.88** | **0.856±0.007** | **0.241±0.004** | **0.907±0.004** |

*This work.